\documentclass[preprint]{elsarticle}
\usepackage{graphicx}
\usepackage{color}
\usepackage{amsmath}
\usepackage{bm}
\usepackage{braket}
\usepackage[colorlinks,linkcolor=blue,citecolor=blue,urlcolor=blue]{hyperref}

\journal{Computer Physics Communications}

\begin{document}

\begin{frontmatter}
\title{Nonrad: Computing Nonradiative Capture Coefficients from First Principles}

\author[1]{Mark E. Turiansky\corref{corauthor}}
\ead{mturiansky@physics.ucsb.edu}
\cortext[corauthor]{Corresponding author.}

\author[2]{Audrius Alkauskas}
\author[3]{Manuel Engel}
\author[3]{Georg Kresse}
\author[4]{Darshana Wickramaratne}
\author[1,5]{Jimmy-Xuan Shen}
\author[6,7]{Cyrus E. Dreyer}
\author[8]{Chris G. Van de Walle}


\address[1]{Department of Physics, University of California, Santa Barbara, CA 93106-9530, U.S.A.}
\address[2]{Center for Physical Sciences and Technology (FTMC), Vilnius LT-10257, Lithuania}
\address[3]{University of Vienna, Faculty of Physics and Center for Computational Materials Sciences, Vienna A-1090, Austria}
\address[4]{Center for Computational Materials Science, US Naval Research Laboratory, Washington DC, 20375, U.S.A.}
\address[5]{Department of Materials Science and Engineering, University of California, Berkeley, CA 94720-1760, U.S.A.}
\address[6]{Department of Physics and Astronomy, Stony Brook University, Stony Brook, New York 11794-3800, U.S.A.}
\address[7]{Center for Computational Quantum Physics, Flatiron Institute, 162 5$^{th}$ Avenue, New York, New York 10010, U.S.A.}
\address[8]{Materials Department, University of California, Santa Barbara, CA 93106-5050, U.S.A.}

\date{\today}

\begin{abstract}
    Point defects in semiconductor crystals provide a means for carriers to recombine nonradiatively.
    This recombination process impacts the performance of devices.
    We present the \texttt{Nonrad} code that implements the first-principles approach of Alkauskas {\it et al.} [Phys. Rev. B \textbf{90}, 075202 (2014)] for the evaluation of nonradiative capture coefficients based on a quantum-mechanical description of the capture process.
    An approach for evaluating electron-phonon coupling within the projector augmented wave formalism is presented.
    We also show that the common procedure of replacing Dirac delta functions with Gaussians can introduce errors into the resulting capture rate, and implement an alternative scheme to properly account for vibrational broadening.
    Lastly, we assess the accuracy of using an analytic approximation to the Sommerfeld parameter by comparing with direct numerical evaluation.
\end{abstract}


\end{frontmatter}

\noindent
{\bf Program summary}

\begin{small}
\noindent
{\em Program Title:\/} Nonrad                                           \\
{\em Program Files doi:\/} \url{https://doi.org/10.5281/zenodo.4274317} \\
{\em Licensing provisions:\/} MIT License                               \\
{\em Programming language:\/} Python 3                                  \\
{\em Nature of problem:\/}
Nonradiative carrier capture at point defects in semiconductors and insulators significantly impacts the performance of devices.
A first-principles approach to calculating nonradiative capture rates is necessary to guide materials design and provide atomistic insight into the nonradiative capture process.
\\
{\em Solution method:\/}
Our code, written in the Python language, implements the first-principles methodology developed by Alkauskas {\it et al.}~[Phys. Rev. B \textbf{90}, 075202 (2014)].
Nonradiative capture rates are calculated using a one-dimensional approach, which maps the prohibitively large phonon problem onto a single, effective phonon mode.
Harmonic phonon matrix elements may then be computed using an analytic technique or by direct numerical integration of the harmonic oscillator wavefunctions.
The electron-phonon coupling is evaluated using the VASP code~[G. Kresse and J. Furthm\"{u}ller, Phys. Rev. B \textbf{54}, 11169 (1996)] to linear order for the effective mode.
\\
\end{small}

\section{Introduction}
\label{sec:intro}
In semiconducting or insulating material, any injected or excited carriers that do not leave the material eventually decay through radiative or nonradiative recombination.
Point defects provide a path for nonradiative transitions, with important implications for device performance.
(In this paper, we use ``point defects'' as a generic term that covers both intrinsic point defects and impurities.)
In optoelectronic devices, point defects allow carriers to recombine nonradiatively through the Shockley-Read-Hall (SRH) process~\cite{abakumov1991nonradiative}.
SRH recombination limits the efficiency of light-emitting diodes, lasers, and photovoltaic cells, transferring the excitation energy into lattice vibrations or, in other words, heat.
Similarly, point defects may act as charge traps or recombination centers in electronic devices such as transistors, degrading performance.
Control of nonradiative recombination is of paramount importance for enhancing the performance of devices.

Many authors have contributed to the theoretical foundations of nonradiative processes in solids~\cite{huang_kun_theory_1950,kubo_application_1955,passler_calculation_1975,henry_nonradiative_1977,stoneham_non-radiative_1981}, dating back several decades.
The work of Henry and Lang~\cite{henry_nonradiative_1977} was particularly influential; it used a semi-classical description and empirical parameters to ascertain the temperature dependence of the nonradiative rates.
While these early works were limited in predictive power, they provided the foundations for a full treatment of the capture process.
Indeed, their limitation was largely due to the fact that they did not take into account the specific electronic and vibronic structure of a given point defect.

First-principles calculations based on state-of-the-art, hybrid density functional theory provide an accurate description of
the electronic structure of point defects~\cite{freysoldt_first-principles_2014}
and a rigorous framework for incorporating atomistic insight into the nonradiative capture process.
A formalism for computing nonradiative capture rates from first principles was developed by Alkauskas {\it et al.}~\cite{alkauskas_first-principles_2014}:
nonradiative recombination rates are computed quantum-mechanically using Fermi's golden rule within the static-coupling approximation.
While the formalism is general, the full multidimensional treatment is computationally extremely expensive.
Alkauskas \textit{et al.} demonstrated that a one-dimensional approximation based on a single phonon mode (known as the accepting mode~\cite{stoneham_non-radiative_1981}) yields very good results, particularly for defects with strong electron-phonon coupling.
Good agreement was demonstrated both with experiment~\cite{alkauskas_first-principles_2014,wickramaratne_iron_2016} and with multidimensional calculations~\cite{wickramaratne_comment_2018,shi_comparative_2015}.

Here, we present \texttt{Nonrad}~\cite{nonrad}, an open-source Python implementation for the computation of nonradiative capture coefficients based on the formalism of Alkauskas \textit{et al}.
\texttt{Nonrad} relies on standard Python libraries such as NumPy~\cite{oliphant_guide_2006} and SciPy~\cite{jones_scipy:_2001} to perform calculations of the nonradiative capture coefficients, as well as Pymatgen~\cite{ong_python_2013} to interface with common first-principles codes.
In addition, \texttt{Nonrad} provides various utilities for preparing and parsing configuration coordinate diagrams.
These utilities facilitate the process of generating input files for these calculations.
Phonon integrals are evaluated using an analytic formula for the overlap between two displaced harmonic oscillator wavefunctions or by direct numerical integration of the analytic wavefunctions.

In addition to the implementation of the approach described in Ref.~\citenum{alkauskas_first-principles_2014}, the present work includes
three important extensions of the methodology:
(i) a method for evaluating electron-phonon matrix elements within the projector augmented wave (PAW) formalism~\cite{blochl_projector_1994},
(ii) an interpolation scheme to obtain a smooth spectral function,
and (iii) numerical evaluation of the Sommerfeld parameters.
The interpolation scheme of extension (ii) is an alternative to replacing the Dirac delta functions with Gaussians, which mimics broadening effects on the energy conservation condition; our scheme provides consistent lineshapes without any tuning parameters.
In extension (iii), we remove the approximation associated with using an analytic form for the Sommerfeld parameter.  We show that this approximation is valid only over a limited range of temperatures, and instead numerically evaluate the integral form of the Sommerfeld parameter.
In total, \texttt{Nonrad} provides a powerful and easy-to-use interface for the evaluation of nonradiative capture coefficients and will enable researchers to reliably investigate nonradiative capture in a range of materials.

\section{Implementation}
\label{sec:implementation}

\subsection{Basic Theory}
\label{sec:basic_theory}

\begin{figure}[!htb]
    \centering
    \includegraphics[width=0.9\columnwidth,height=0.3\textheight,keepaspectratio]{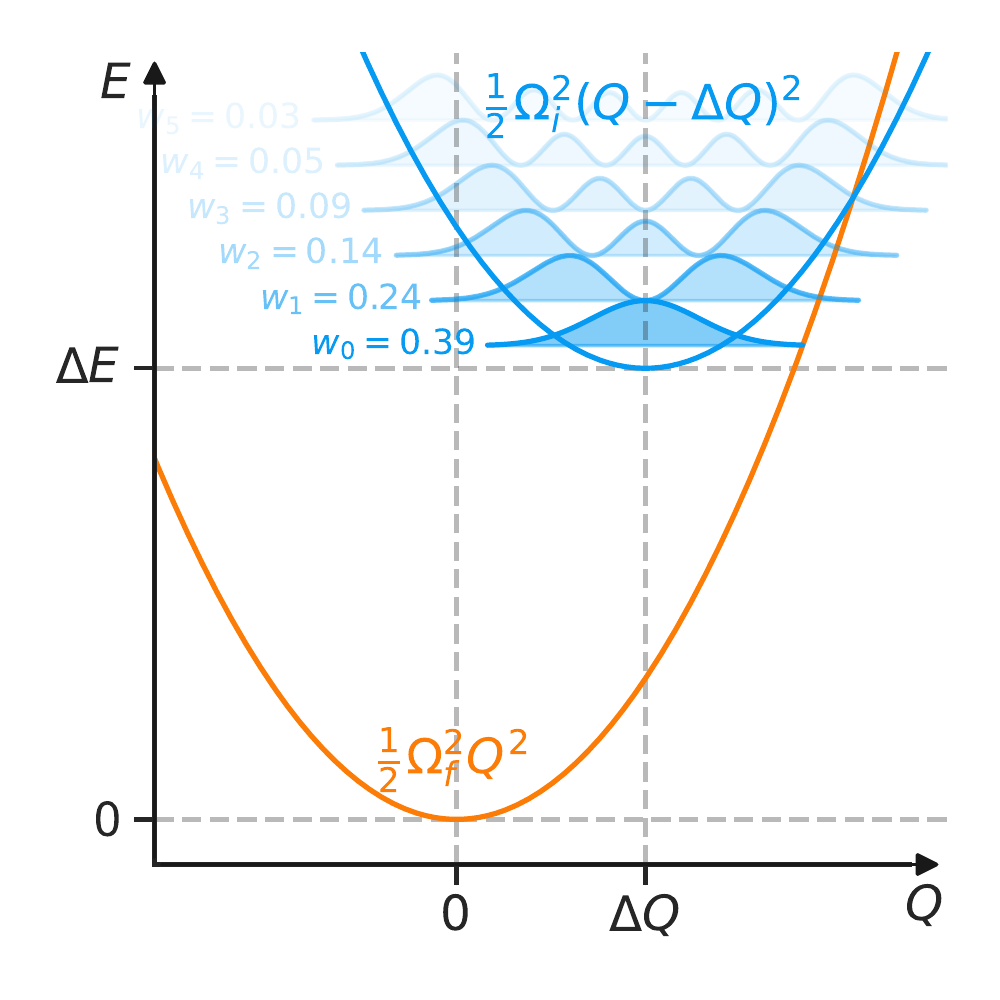}
    \caption{\label{fig:schematic}
        A schematic configuration coordinate diagram.
        The blue parabola corresponds to the initial state with a phonon frequency $\Omega_i$ centered at the equilibrium geometry, defined to be $Q = \Delta Q$.
        The orange parabola corresponds to the final state with frequency $\Omega_f$ and equilibrium geometry $Q = 0$.
        For the initial state, harmonic oscillator probability densities are shown, offset by the energy of the state.
        The opacity of the probability densities correspond to the thermal occupation of each state $w_m$ at a temperature $k_B T = 2 \hbar \Omega_i$.
    }
\end{figure}

Calculations of nonradiative capture begin with accurate modeling of the point defect using the standard methodology~\cite{freysoldt_first-principles_2014} in which charged point defects are studied within a supercell and finite-size corrections are included~\cite{freysoldt_fully_2009}.
The nonradiative capture coefficient $C$ describes capture of a carrier at the band edge onto the defect.
The calculation of $C$ in \texttt{Nonrad} entails the evaluation of eq~22 from Ref.~\citenum{alkauskas_first-principles_2014}, reproduced here:
\begin{equation}
    \label{eq:cap_coeff}
    C = f \frac{2 \pi}{\hbar} g V W_{if}^2 \sum_m w_m \sum_n {\lvert \braket{\chi_{im} \lvert \hat{Q} - Q_0 \rvert \chi_{fn}} \rvert}^2 \delta (\Delta E + m\hbar\Omega_i - n\hbar\Omega_f) \;.
\end{equation}
Here, $g$ is the configurational degeneracy: multiple atomic configurations may exist with degenerate (or nearly degenerate) energies.
The summations in Eq.~\ref{eq:cap_coeff} are over the quantum numbers of the special phonon mode known as the accepting mode~\cite{stoneham_non-radiative_1981}.
This mode corresponds to the mass-weighted configuration coordinate $Q$ used to construct the configuration coordinate diagram in Fig.~\ref{fig:schematic}.
We define $\Delta Q$ as the difference in the equilibrium geometries of the initial and final state along this configuration coordinate.
Such diagrams describe the change in energy as the atomic coordinates change for a given charge state of a defect.
$\Delta E$ is the energy difference between the initial and final state, $\Omega_{\{i,f\}}$ are the harmonic phonon frequencies of the initial and final state, and $w_m$ is the occupation factor in the Bose-Einstein distribution for the initial state.
In Eq.~\ref{eq:cap_coeff}, $Q_0$ is the atomic configuration used as the starting point for the perturbative expansion.  This can be chosen to correspond to the equilibrium geometry of either the initial or the final state, as will be discussed in Sec.~\ref{sec:elph}.

In this work, we focus on the case where the initial state corresponds to a delocalized band-edge state. In this case, $V$ is the volume of the supercell used for the simulation. The \texttt{Nonrad} code, however, is equally applicable to the case where the initial state is a localized excited state of the defect.
In that case, $V$ and the scaling function $f$ (discussed below) should both be set equal to 1, and a rate is obtained (in units of s$^{-1}$) as opposed to a capture coefficient (in units of cm$^3$s$^{-1}$).

The phonon matrix elements $\braket{\chi_{im} \lvert \hat{Q} - Q_0 \rvert \chi_{fn}}$ are evaluated within the \texttt{Nonrad} code by writing them in terms of wavefunction overlaps through the ladder operators $\hat{Q} = \sqrt{\hbar / 2 \Omega} (\hat{a}^\dagger + \hat{a})$, where $\hat{a}^\dagger$ and $\hat{a}$ are the raising and lowering operators.
The resulting overlaps are calculated using either the analytic formula of Zapol~\cite{zapol_new_1982} or by direct numerical integration of the harmonic oscillator wavefunctions, which is the default.
These two approaches produce numerically identical results, but differ in their computational cost.
For small vibronic quantum numbers, the analytic formula is faster, while numerical integration becomes faster in the opposite limit.

The scaling function $f$ in Eq.~\ref{eq:cap_coeff} accounts for two contributions: (i) the Coulombic interaction between the defect and delocalized carrier if the defect is charged and
(ii) the perturbation of the band-edge wavefunction due to the interaction with the charged defect level in a finite-size supercell.
In previous works, a more general form of the scaling function was defined [see eq~14 in Ref.~\citenum{alkauskas_first-principles_2014}];
the more general form reduces to those described below for typical simulation conditions.
Contribution (i) has been described in the literature and is known as the Sommerfeld parameter~\cite{passler_relationships_1976}.
The Sommerfeld parameter may be calculated using the \verb|sommerfeld_parameter| function available in the \texttt{Nonrad} code.
We examine the accuracy of the commonly used analytic form of the Sommerfeld parameter in Sec.~\ref{subsec:sp}.
The effect of contribution (ii) can be estimated by comparing the band-edge wavefunction in the supercell to a homogeneous distribution, where the integrated radial charge density goes as $(4/3) \pi R^3 / V$.
If we define $\alpha$ as the ratio of the integrated radial distribution of the band-edge charge density to the homogeneous case, then the resulting scaling of the capture coefficient is $\alpha^{-1}$.
The procedure for calculating this contribution is encapsulated in the \verb|charged_supercell_scaling| function in the \texttt{Nonrad} code.
The final scaling function is the product of contributions (i) and (ii), and we set it to 1 for capture of a carrier in the neutral charge state with the electron-phonon matrix element evaluated in a neutral cell.

The electron-phonon matrix element $W_{if}$ needs to be computed from first principles.
The PAW pseudopotential formalism has become a prominent framework for computation because it provides a large reduction in computational complexity while still retaining an effective full-potential wavefunction, albeit with the core electrons frozen.
To compute matrix elements within the PAW formalism, the contribution from the overlap operator needs to be accounted for.
In Sec.~\ref{subsec:paw}, we present a linear response theory for evaluating the electron-phonon matrix elements within the PAW formalism.
Finally, we examine the common practice of replacing the energy-conserving delta functions in Eq.~\ref{eq:cap_coeff} with Gaussians and show that it may introduce significant errors in the resulting capture coefficients.
We propose an alternative scheme in Sec.~\ref{subsec:interp}.

\subsection{PAW Formalism}
\label{subsec:paw}

We begin by writing the matrix element in terms of the pseudo-wavefunctions:
\begin{equation}
    \label{eq:wif}
    W_{if} = \braket{\psi_i \lvert \partial_Q \hat{H} \rvert \psi_f}
    \approx \braket{{\tilde \psi}_i \lvert \partial_Q \hat{\tilde H} - \epsilon_f \partial_Q \hat{\tilde S} \rvert {\tilde \psi}_f} \;,
\end{equation}
where $\psi_j$ is the single-particle wavefunction and $\hat{H}$ is the full-potential Kohn-Sham Hamiltonian.
$\hat{\tilde H}$ is the Kohn-Sham Hamiltonian for the pseudo-wavefunctions after applying the PAW transformation, $\hat{\tilde S}$ is the overlap matrix arising from the non-orthogonality of the pseudo-wavefunctions, and $\epsilon_j$ is the energy eigenvalue corresponding to the pseudo-wavefunction ${\tilde \psi}_j$.
The right-hand side of the equation is obtained by
first performing the PAW transformation ($\hat{\mathcal T} {\tilde \psi}_j={\psi}_j$), then using the PAW completeness relation, and finally taking derivatives.
This results in an approximation, where the term $\braket{\tilde{\psi}_i \lvert \hat{\mathcal T}^\dagger \partial_Q \hat{\mathcal T} \rvert \tilde{\psi}_f}$ is assumed to be negligible.
Chaput and coworkers~\cite{chaput2019} recently suggested an alternative approach, which first takes the derivative on the left hand side, then performs the PAW transformation, and finally uses the completeness relation.
The relation between these two approaches is discussed elsewhere \cite{engel}.

Equation~\ref{eq:wif} is unwieldy to evaluate directly, and a more tractable form may be derived from linear response theory.
The effective Kohn-Sham equation for the pseudo-wavefunctions,
\begin{equation}
    \label{eq:ks}
    (\hat{\tilde H} - \epsilon_f \hat{\tilde S}) \ket{{\tilde \psi}_f} = 0 \;,
\end{equation}
is expanded to linear order in the perturbation:
\begin{equation}
    \label{eq:expansion}
    \left[ (\hat{\tilde H} + \partial_Q \hat{\tilde H}) - (\epsilon_f + \partial_Q \epsilon_f)(\hat{\tilde S} + \partial_Q \hat{\tilde S}) \right]
    \ket{{\tilde \psi}_f + \partial_Q {\tilde \psi}_f} = 0 \;.
\end{equation}
Collecting terms that are linear in the perturbation and equating them to 0 gives
\begin{equation}
    \label{eq:collected}
    (\hat{\tilde H} - \epsilon_f \hat{\tilde S}) \ket{\partial_Q {\tilde \psi}_f} =
    - (\partial_Q \hat{\tilde H} - \epsilon_f \partial_Q \hat{\tilde S}) \ket{{\tilde \psi}_f} + (\partial_Q \epsilon_f) \hat{\tilde S} \ket{{\tilde \psi}_f} \;.
\end{equation}
Finally, we multiply by $\bra{{\tilde \psi}_i}$, obtaining
\begin{equation}
    \label{eq:multiply0}
    \braket{{\tilde \psi}_i \lvert \hat{\tilde H} - \epsilon_f \hat{\tilde S} \rvert \partial_Q {\tilde \psi}_f} =
    - \braket{{\tilde \psi}_i \lvert \partial_Q \hat{\tilde H} - \epsilon_f \partial_Q \hat{\tilde S} \rvert {\tilde \psi}_f} +
    (\partial_Q \epsilon_f) \braket{{\tilde \psi}_i \lvert \hat{\tilde S} \rvert {\tilde \psi}_f}
\end{equation}
or
\begin{equation}
    \label{eq:multiply1}
    (\epsilon_f - \epsilon_i) \braket{{\tilde \psi}_i \lvert \hat{\tilde S} \rvert \partial_Q {\tilde \psi}_f} =
    \braket{{\tilde \psi}_i \lvert \partial_Q \hat{\tilde H} - \epsilon_f \partial_Q \hat{\tilde S} \rvert {\tilde \psi}_f} \;,
\end{equation}
where we have used Eq.~\ref{eq:ks} and the orthogonality relation $\braket{{\tilde \psi}_i \lvert \hat{\tilde S} \rvert {\tilde \psi}_j} = \delta_{ij}$.
We can now plug this into Eq.~\ref{eq:wif} to arrive at the final form for the electron-phonon matrix element:
\begin{equation}
    \label{eq:pawwif}
    W_{if} = (\epsilon_f - \epsilon_i) \braket{{\tilde \psi}_i \lvert \hat{\tilde S} \rvert \partial_Q {\tilde \psi}_f}\;.
\end{equation}
In practice, Eq.~\ref{eq:pawwif} is evaluated by computing the slope of $\braket{{\tilde \psi}_i (0) \lvert \hat{\tilde S} (0) \rvert {\tilde \psi}_f (Q)}$ as a function of $Q$.

Other methods for evaluating the electron-phonon coupling within the PAW formalism have been described in the literature, for instance in Refs.~\citenum{chaput2019} and \citenum{engel} and in Appendix C of Ref.~\citenum{barmparis_theory_2015};
however, they rely on computing structure factors resulting from the displacement of the PAW spheres or implementing the more involved Eq.~\ref{eq:wif}.
The method described here is straightforward, very efficient and has been implemented in the VASP code version 5.4.4, as well as version 6.

\subsection{Broadening of the Delta Functions}
\label{subsec:interp}
The delta functions in Eq.~\ref{eq:cap_coeff} impose energy conservation on the vibronic transitions.
In reality, the delta functions are too restrictive: various mechanisms act to broaden the energies of the transitions, producing a continuous spectral function.
In solids the dominant sources of broadening are random internal fields and the finite lifetime of the states involved in the transition, which leads to energy uncertainty.~\cite{stoneham_theory_1975}
To simulate this broadening, the delta functions are typically replaced with Gaussians, whose width is characterized by a smearing parameter.
Within the one-dimensional approximation, the broadening is relatively large because it needs to reflect the contributions from many phonon modes; unfortunately, this renders
the results highly sensitive to the choice of the smearing parameter.
This effect was examined in the case of luminescence lineshapes within the single-phonon-mode approximation~\cite{alkauskas_tutorial:_2016}.
In the context of nonradiative capture, this effect can be understood by examining the vibrational lineshape function, defined as~\cite{stoneham_non-radiative_1981}:
\begin{equation}
    \label{eq:lineshape}
    G_m (\omega) = \sum_n {\lvert \braket{\chi_{im} \lvert \hat{Q} - Q_0 \rvert \chi_{fn}} \rvert}^2 \delta (\omega + m\Omega_i - n\Omega_f) \;.
\end{equation}

\begin{figure}[!htb]
    \centering
    \includegraphics[width=0.9\columnwidth,height=0.5\textheight,keepaspectratio]{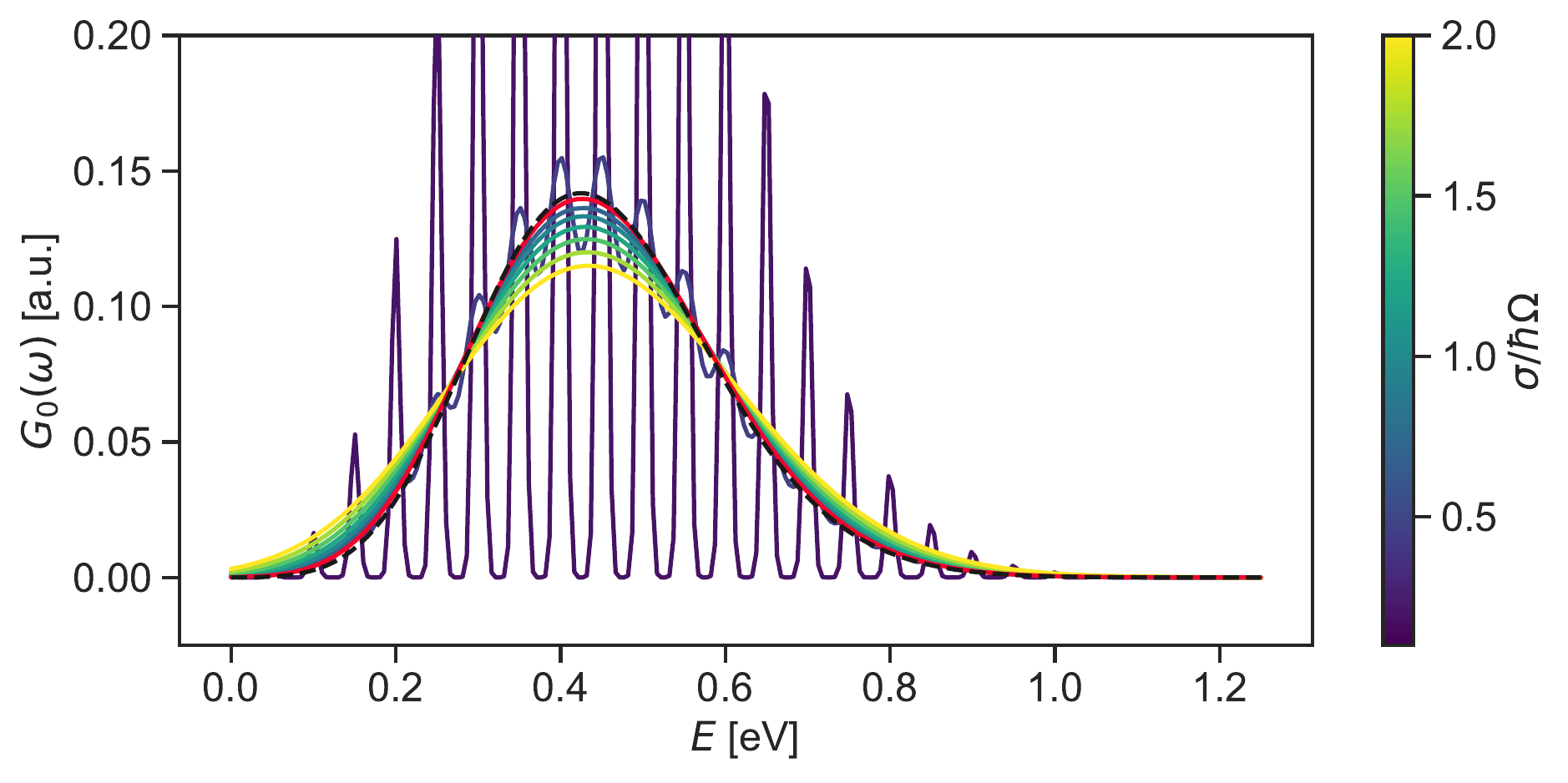}
    \caption{\label{fig:lineshape}
        The lineshape function $G_0$ for $\hbar\Omega = 0.05$~eV and with Huang-Rhys factor $S = 8$.
        In the figure, the black, dashed line comes from the analytic form of the lineshape function (Eq.~\ref{eq:pekar_cont}) from the Pekarian function~\cite{di_bartolo_advances_1991}.
        The red line corresponds to the lineshape function resulting from our proposed interpolation scheme.
        The remaining lines are generated using Gaussians for the delta functions, with the color corresponding to the smearing parameter $\sigma$.
        If the smearing parameter is too small, the lineshape function is not sufficiently smooth.
        For too large a smearing parameter, the tails of the Gaussians add up, which can be seen in the low- and high-energy portions of the lineshape function.
    }
\end{figure}

When we replace the delta function in Eq.~\ref{eq:lineshape} with a Gaussian, we are left with conflicting requirements for the choice of the smearing parameter $\sigma$.
On the one hand, if $\sigma$ is too small, the lineshape function will not be smooth and will perform barely better than the delta functions.
Too large a value for $\sigma$, on the other hand, results in the ``tails'' of the Gaussians from various transitions adding up and creating an artificial enhancement of the function at high and low energies.
These effects are illustrated in Fig.~\ref{fig:lineshape}, where the lineshape function for the lowest transition $G_0 (\omega)$ is shown for various choices of $\sigma$.
In Fig.~\ref{fig:lineshape}, we have assumed that the phonon frequencies are identical in the initial and final state ($\Omega_i = \Omega_f = \Omega$), and that the Huang-Rhys factor $S$, a dimensionless parameter quantifying the electron-phonon coupling strength, is an integer.
Under these assumptions, the lineshape function can be obtained from the Pekarian function~\cite{di_bartolo_advances_1991}:
\begin{equation}
    \label{eq:pekar}
    G_0 (\omega) = \sum_n \frac{e^{-S} S^n}{n!} \delta (\omega - n\Omega) \;.
\end{equation}
To obtain a continuous function, we treat $n$ as a continuous variable, replace the summation over $n$ with an integral, and integrate to obtain
\begin{equation}
    \label{eq:pekar_cont}
    G_0 (\omega) = \frac{e^{-S} S^{\omega/\Omega}}{\Gamma (\omega/\Omega + 1)}\;.
\end{equation}
This function is shown for comparison in Fig.~\ref{fig:lineshape}.

Our solution is to use cubic spline interpolation to produce a smooth lineshape function, $g_m (\omega)$.
In practice, the data points given by $\left( n\Omega_f-m\Omega_i, {\lvert \braket{\chi_{im} \lvert \hat{Q} - Q_0 \rvert \chi_{fn}} \rvert}^2 \right)$ are interpolated to produce the function $g_m (\omega)$.
The use of cubic splines ensures the smoothness of $g_m$ and removes the need for a smearing parameter.
To ensure consistency between the interpolated function and the true lineshape, we normalize $g_m$ such that
\begin{equation}
    \label{eq:lineshape_norm}
    \int d\omega\, g_m (\omega) = \int d\omega\, G_m (\omega) = \sum_n {\lvert \braket{\chi_{im} \lvert \hat{Q} - Q_0 \rvert \chi_{fn}} \rvert}^2\;.
\end{equation}
The proposed interpolation scheme provides a reliable representation of the lineshape function, reproducing Eq.~\ref{eq:pekar_cont} for the conditions stated above (see Fig.~\ref{fig:lineshape}).

\begin{figure}[!htb]
    \centering
    \includegraphics[width=0.9\columnwidth,height=0.3\textheight,keepaspectratio]{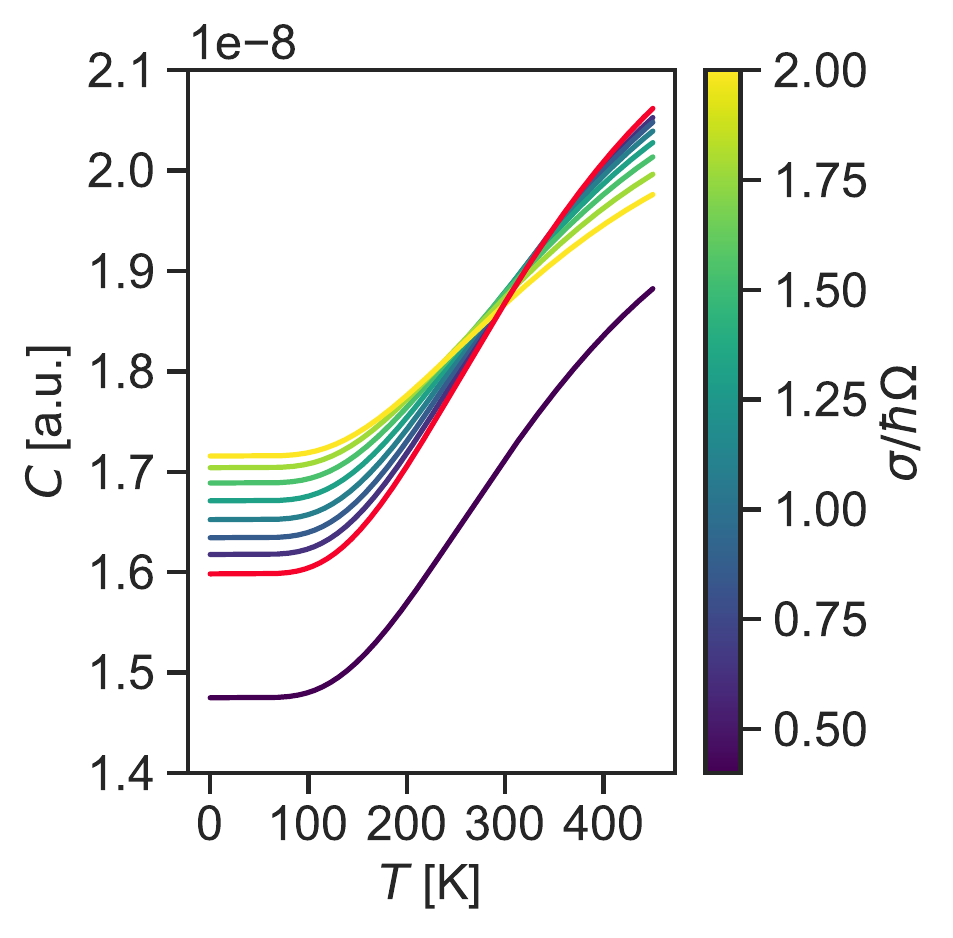}
    \caption{\label{fig:comparison}
        Capture coefficient for the case of $\hbar\Omega = 0.05$~eV and with Huang-Rhys factor $S = 8$.
        The red line corresponds to the capture coefficient determined using the proposed interpolation scheme.
        The remaining lines are generated using Gaussians to replace the delta functions in Eq.~\ref{eq:cap_coeff},
        with the color corresponding to the smearing parameter $\sigma$.
        Large deviations are seen for the Gaussian scheme.
    }
\end{figure}

The method for treating the delta functions impacts not only the lineshape function but also the nonradiative capture coefficient.
A comparison of the capture coefficient obtained with Gaussian smearing and with the new interpolation scheme is shown in Fig.~\ref{fig:comparison}.
From the figure, it is clear that use of the interpolation scheme prevents erroneous values in the capture coefficient.

\subsection{Sommerfeld Parameter}
\label{subsec:sp}
The Sommerfeld parameter describes the enhancement or suppression of a delocalized wavefunction near a charge.
When a carrier is captured by a charged defect, the Sommerfeld parameter must be included to scale the capture coefficient correctly.
Typically, an analytic form for the Sommerfeld parameter is used, given by~\cite{passler_relationships_1976}:
\begin{equation}
    \label{eq:sp}
    s(T) = \begin{cases}
        \frac{4}{\sqrt{\pi}} {\left[ \frac{Z^2 \theta_b}{T} \right]}^{1/2} & Z < 0 \\
        \frac{8}{\sqrt{3}} {\left[ \frac{Z^2 \theta_b}{T} \right]}^{2/3} \exp \left( -3 {\left[ \frac{Z^2 \theta_b}{T} \right]}^{1/3} \right) & Z > 0 \\
    \end{cases}\;,
\end{equation}
where $\theta_b = m_b e^4 / 32 k_B \epsilon_0^2 \hbar^2$ is a temperature parameter, $e$ is the fundamental charge, $m_b$ is the band effective mass, $\epsilon_0$ is the static dielectric constant, $k_B$ is the Boltzmann constant, and $Z$ is the ratio of the defect charge to the carrier charge ($Z > 0$ corresponds to a repulsive center).

In deriving Eq.~\ref{eq:sp}, the Sommerfeld parameter is first obtained as a function of momentum and written as
\begin{equation}
    \label{eq:spk}
    s({\bm k}) = -\frac{2 \pi Z}{a_b \lvert {\bm k} \rvert} \frac{1}{1 - e^{2 \pi Z / a_b \lvert {\bm k} \rvert}}\;,
\end{equation}
where $a_b = 4 \pi \epsilon_0 \hbar^2 / m_b e^2$ is an effective Bohr radius.
Temperature averaging is then performed to give
\begin{equation}
    \label{eq:spT}
    s(T) = \frac{\int_0^\infty d \lvert {\bm k} \rvert \, 4 \pi {\lvert {\bm k} \rvert}^2 \, s({\bm k}) e^{-\hbar^2 {\lvert {\bm k} \rvert}^2 / 2 m_b k_B T}}{\int_0^\infty d \lvert {\bm k} \rvert \, 4 \pi {\lvert {\bm k} \rvert}^2 \, e^{-\hbar^2 {\lvert {\bm k} \rvert}^2 / 2 m_b k_B T}} \;.
\end{equation}
At this point, an approximation is necessary to obtain the analytic form given in Eq.~\ref{eq:sp}.
It is assumed that $\lvert {\bm k} \rvert \ll 2 \pi \lvert Z \rvert / a_b$, allowing to write $s ({\bm k})$ in Eq.~\ref{eq:spk} as
\begin{equation}
    \label{eq:spk_approx}
    s({\bm k}) \approx \frac{2 \pi \lvert Z \rvert}{a_b \lvert {\bm k} \rvert} \begin{cases}
        1 & Z < 0 \\
        e^{-2 \pi Z / a_b \lvert {\bm k} \rvert} & Z > 0
    \end{cases}\;.
\end{equation}
The integrals in Eq.~\ref{eq:spT} may then be evaluated analytically to give the form of $s(T)$ in Eq.~\ref{eq:sp}.

\begin{figure}[!htb]
    \centering
    \includegraphics[width=0.9\columnwidth,height=0.5\textheight,keepaspectratio]{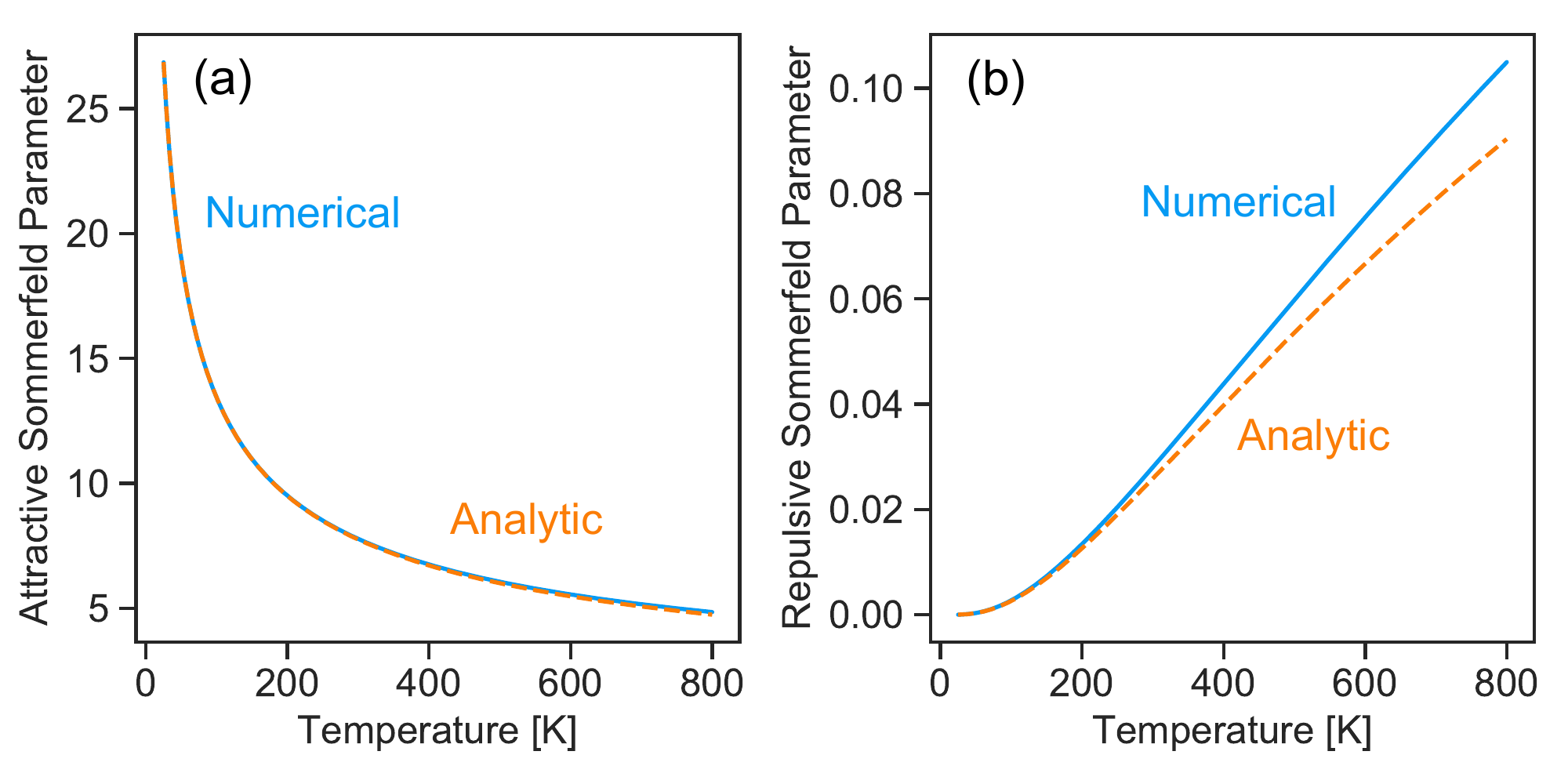}
    \caption{\label{fig:sp}
        The Sommerfeld parameter in the (a) attractive and (b) repulsive cases with the parameters of GaN ($m_b = 0.18$ and $\epsilon_0 = 8.9$).
        Blue solid lines correspond to the numerical integration of Eq.~\ref{eq:spT}, and orange dashed lines correspond to the analytic form of Eq.~\ref{eq:sp}.
        Significant deviations are apparent for the repulsive case.
    }
\end{figure}

While the analytic form is simple to work with, the assumptions made in its derivation may not be valid for the temperatures where it is applied.
Instead of making the approximation above, Eq.~\ref{eq:spT} can be evaluated numerically to give a more exact value for the Sommerfeld parameter.
A comparison of the analytic form and direct numerical evaluation of the Sommerfeld parameter for the example of electron capture at a defect in GaN is shown in Fig.~\ref{fig:sp}.
In the case of a repulsive interaction, the analytic form leads to errors exceeding 15\% at high temperature;
even at room temperature, an error of 7.7\% is made.
For attractive interactions, the errors are less severe, only surpassing 2\% at high temperatures.
For the implementation in \texttt{Nonrad}, direct numerical integration of the Sommerfeld parameter is utilized, with the analytic form provided for comparison.

\section{Numerical Example}
\label{sec:tutorial}

In this section, we provide an example calculation of the nonradiative capture rate for a well-studied defect system: a carbon atom substituting on a nitrogen site (C$_{\rm N}$) in GaN.
We show how the utilities provided by \texttt{Nonrad} can simplify the procedure.
This information is also available as a Jupyter notebook~\cite{kluyver_jupyter_2016} distributed with the \texttt{Nonrad} code.\footnote{Accessible at \url{https://github.com/mturiansky/nonrad/blob/master/notebooks/tutorial.ipynb}}
While the example here is performed with the VASP code~\cite{kresse_efficient_1996}, we emphasize that the final evaluation of the capture coefficient relies only on the derived parameters that may be obtained from any first-principles code
that implements a method to obtain the electron-phonon coupling.

\subsection{Computational Details}
\label{sec:comp_det}

Calculations of the capture coefficient are sensitive to the quality of the input parameters; therefore, highly accurate first-principles methods should be employed.
In this work, we utilize density functional theory as implemented in version 5.4.4 of the VASP code~\cite{kresse_efficient_1996} with the hybrid functional of Heyd, Scuseria, and Ernzerhof (HSE)~\cite{heyd_erratum:_2006}.
The HSE mixing parameter of 0.31 is chosen to yield a band gap ($E_g = 3.54$ eV) and lattice parameters ($a = 3.20$ {\AA} and $c = 5.19$ {\AA}) in agreement with the experimental values~\cite{schulz_crystal_1977,vurgaftman_band_2003}.
Calculations are performed within the PAW formalism~\cite{blochl_projector_1994} with an energy cutoff of 400 eV.
Ga $d$ states are included in the core.
The atomic configurations are relaxed until forces are below 5 meV/{\AA}.
A 96-atom supercell is used to simulate the defect:
the 4-atom primitive cell is transformed to an 8-atom orthorhombic cell, and then a $3\times2\times2$ multiple of this orthorhombic cell produces the 96-atom supercell.
A single, special \textbf{k} point is used for the Brillouin-zone sampling~\cite{baldereschi_mean-value_1973} to obtain the atomic structure.
With these choices, we find that in the neutral charge state the structure is distorted along one of the planar bonds;
the C-Ga bond lengths are {1.97 \AA} (axial) and {2.03 \AA}, {2.00 \AA}, and {1.98 \AA}, compared to the N-Ga bulk values of {1.96 \AA} (axial) and {1.95 \AA} (planar).
In the negative charge state, atomic displacements correspond to a breathing relaxation, with C-Ga bond lengths of {1.94 \AA} (axial) and {1.93 \AA} (planar).

Electron-phonon matrix elements are evaluated at the $\Gamma$ point.

\subsection{Energetics}
\label{sec:energetics}

\begin{figure}[!htb]
    \centering
    \includegraphics[width=0.9\columnwidth,height=0.5\textheight,keepaspectratio]{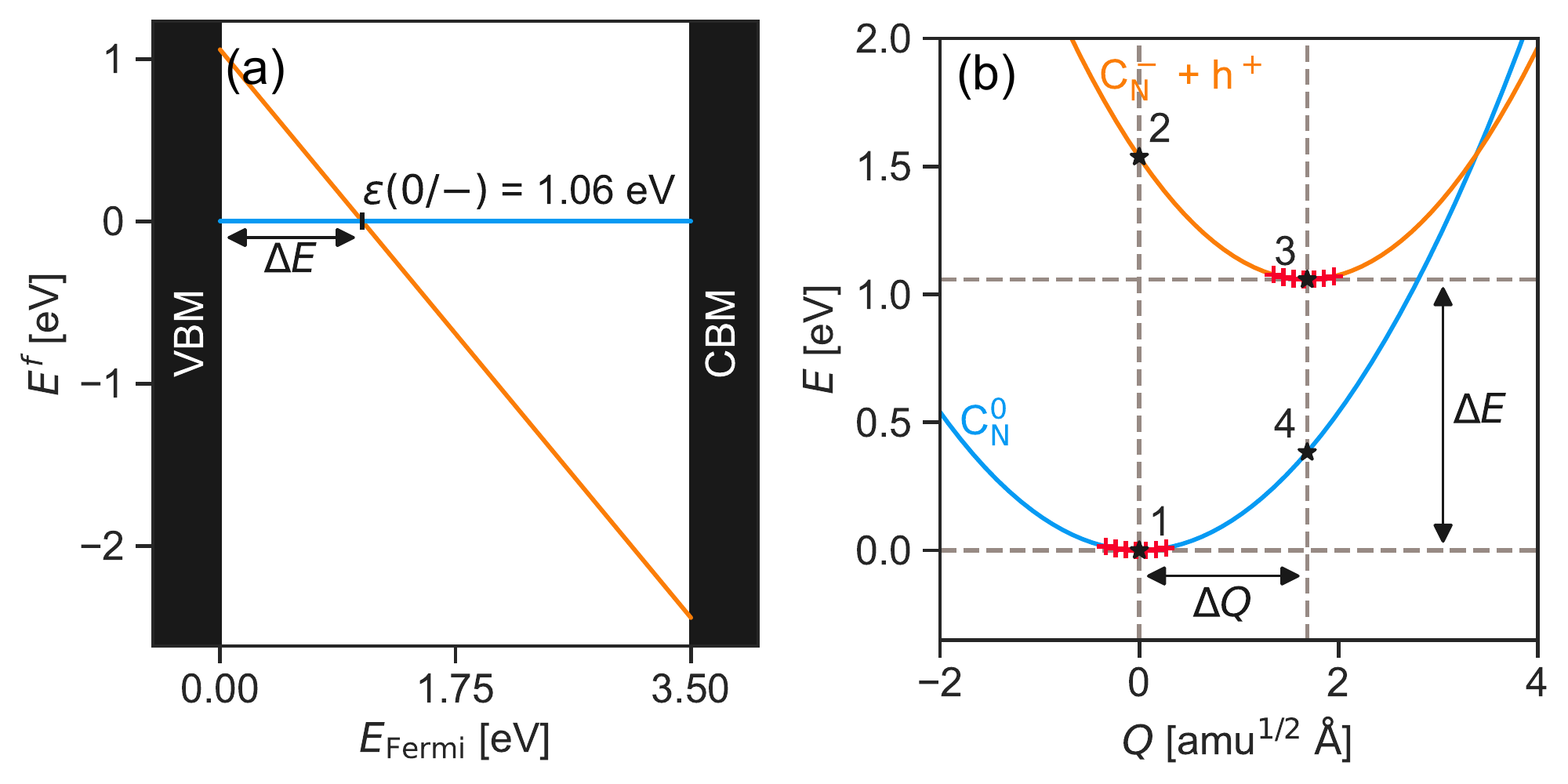}
    \caption{\label{fig:form_en_ccd}
        (a) Formation energy diagram and (b) configuration coordinate diagram for the C$_{\rm N}$ defect system.
        For the purposes of the plot, the formation energy of the neutral charge state is set to zero.
        In both figures, the blue line corresponds to the neutral charge state and the orange line to the negative charge state.
        The red crosses in (b) correspond to actual calculations at interpolated geometries.
        The coordinate corresponding to the equilibrium geometry of the final state has been set to zero.
        Points 1-4 in (b) label possible configurations for computing the electron-phonon coupling.
    }
\end{figure}

We begin with a calculation of the formation energy of C$_{\rm N}$ in GaN within the point-defect formalism of Ref.~\citenum{freysoldt_first-principles_2014}.
The formation energy for defect $X$ in charge state $q$ is given by
\begin{equation}
    \label{eq:form_en}
    E^f [X^q] = E_{\rm tot} [X^q] - E_{\rm tot} [{\rm Bulk}] - \sum_i n_i \mu_i + q (E_{\rm VBM} + E_F) + \Delta_q \;,
\end{equation}
where $E_{\rm tot} [X^q]$ is the total energy of the supercell containing the defect and $E_{\rm tot} [{\rm Bulk}]$ is the total energy of the pristine supercell.
$E_{\rm VBM}$ is the energy of the valence-band maximum, $E_F$ is the Fermi level, and $\Delta_q$ is a correction term that accounts for the effects that arise from the use of a finite-size supercell~\cite{freysoldt_fully_2009,freysoldt_electrostatic_2011}.
The chemical potential $\mu_i$ accounts for the addition ($n_i > 0$) or removal ($n_i < 0$) of species in forming the defect.
While the chemical potentials are necessary to determine relative stability of different types of defects, we are interested in the electronic properties of a single defect here.  Specifically, carrier capture depends on the thermodynamic transition level, which
is determined by the Fermi-level position where a change of charge state occurs; it is defined as
\begin{equation}
    \label{eq:eftl}
    \varepsilon(q/q^\prime) = \frac{E^f [X^q; E_F = 0] - E^f [X^{q^\prime}; E_F = 0]}{q^\prime - q}\;,
\end{equation}
where $E^f [X^q; E_F = 0]$ is the formation energy at the valence-band maximum ($E_F = 0$).
Using Eq.~\ref{eq:form_en}, Eq.~\ref{eq:eftl} can be rewritten as
\begin{equation}
    \label{eq:tl}
    \varepsilon(q/q^\prime) = \frac{(E_{\rm{tot}} [X^q] + \Delta_q) - (E_{\rm{tot}} [X^{q^\prime}] + \Delta_{q^\prime})}{q^\prime - q} - E_{\rm VBM}\;.
\end{equation}
The formation energy and transition level for the C$_{\rm N}$ defect are shown in Fig.~\ref{fig:form_en_ccd}a.
We find a $\varepsilon(0/-)$ value of 1.06 eV; this value differs slightly from previous work (1.02 eV) due to minor differences in the finite-size correction~\cite{alkauskas_first-principles_2014}.

In this tutorial example, we will calculate the hole capture rate; for hole capture, the energy difference between the initial and the final state corresponds to the thermodynamic transition level, i.e., $\Delta E = \varepsilon(0/-)$.
If we were considering electron capture, the energy difference should be taken with respect to the conduction-band minimum, i.e., $\Delta E = E_g - \varepsilon(0/-)$, where $E_g$ is the band gap.
$\Delta E$ is one input parameter to the \texttt{Nonrad} code.

\subsection{Configuration Coordinate Diagram}
\label{sec:ccd}

The configuration coordinate diagram (CCD) in Fig.~\ref{fig:form_en_ccd}b not only
provides a convenient illustration of the nonradiative capture process but also allows us to derive important input parameters for the \texttt{Nonrad} code.
The CCD is prepared by linearly interpolating between the equilibrium structures of the two charge states involved in the nonradiative transition.
These interpolated structures can be generated using the \verb|get_cc_structures| function provided by \texttt{Nonrad}.
The energy of each interpolated structure in the two charge states is plotted as a function of the generalized configuration coordinate $Q$.
As mentioned in Sec.~\ref{sec:basic_theory}, this coordinate $Q$
corresponds to the special phonon mode used in the one-dimensional approximation.
Based on the potential energy surfaces plotted along $Q$, we can extract the phonon frequencies in the initial and final states.

Figure~\ref{fig:form_en_ccd}b for C$_{\rm N}$ was prepared using the utilities provided by \texttt{Nonrad}.
The initial state corresponds to the C$_{\rm N}$ defect in the negative charge state and a hole in the valence band, and the final state corresponds to the defect in the neutral charge state.
From the CCD, we extract the separation between the equilibrium structures along the configuration coordinate $\Delta Q$ in units of $\sqrt{\rm amu}$ {\AA}, using the \verb|get_dQ| function.
Also, we find the phonon frequencies in the initial and final states, $\Omega_{\{i,f\}}$ in eV, using the \verb|get_PES_from_vaspruns| and \verb|get_omega_from_PES| functions.
For this system, we obtain the values $\Delta Q = 1.69$ $\sqrt{\rm amu}$ \AA, $\hbar\Omega_i = 0.0375$ eV, and $\hbar\Omega_f = 0.0336$ eV.

\subsection{Electron-Phonon Coupling}
\label{sec:elph}

Next we compute the electron-phonon coupling using the implementation in VASP\@.
We need to determine the atomic configuration $Q_0$ and charge state of the defect that will be used for computing the electron-phonon matrix elements;
this gives us four points to choose from, labeled 1-4 in Fig.~\ref{fig:form_en_ccd}b.
Points 1 and 2 correspond to the equilibrium geometry of the final state ($Q = 0$), while points 3 and 4 correspond to the geometry of the initial state ($Q = \Delta Q$).
For evaluating the electron-phonon coupling, it's important to identify an equilibrium geometry with a clearly identifiable defect state in the band gap.
For C$_{\rm N}$, a Kohn-Sham state associated with the defect state into which the hole is being captured can be clearly identified in the calculations performed for the geometry of the neutral charge state; therefore the geometry of the final state is a good choice for $Q_0$.
We still need to choose between points 1 and 2;
these points differ in the charge of the supercell.
Point 1 is most convenient to use because we already have the wavefunctions available at displaced geometries from the preparation of the CCD\@.
Furthermore, point 1 corresponds to the neutral charge state and avoids the complications that come with the use of a charged supercell.
If we were to choose the negative charge state (point 2), we would need to include contribution (ii) of the scaling function to account for the interaction between the defect and the band-edge wavefunctions, as discussed in Sec.~\ref{sec:basic_theory}.

To prepare the VASP calculation of the overlaps for the electron-phonon coupling, the \texttt{KPOINTS} and \texttt{POTCAR} files from the calculation of point 1 are used.
The \texttt{INCAR} file should now contain \verb|LWSWQ = True| and \verb|ALGO = None|.
The \texttt{WAVECAR} files for each of the displaced geometries around point 1 are then copied to \texttt{WAVECAR.qqq} and
VASP is executed once for each available \texttt{WAVECAR.qqq} file (the  \texttt{POSCAR} and \texttt{WAVECAR} files
must correspond to the undisplaced geometries at point 1, and VASP will read both \texttt{WAVECAR} files to calculate
$\braket{{\tilde \psi}_i (0) \lvert \hat{\tilde S} (0) \rvert {\tilde \psi}_f (Q)}$).
These calculations will prepare a series of \texttt{WSWQ} files, one for each displacement, which the \verb|get_Wif_from_WSWQ| function provided by \texttt{Nonrad} can process.
In GaN, the VBM contains three (nearly degenerate) bands; we mean-square average over the electron-phonon matrix elements with these bands.
This procedure yields the electron-phonon matrix element $W_{if} = 0.050$ eV amu$^{-1/2}$ \AA$^{-1}$ for the C$_{\rm N}$ defect.

\subsection{Capture Coefficient}
\label{sec:cap_coeff}

\begin{figure}[!htb]
    \centering
    \includegraphics[width=0.9\columnwidth,height=0.5\textheight,keepaspectratio]{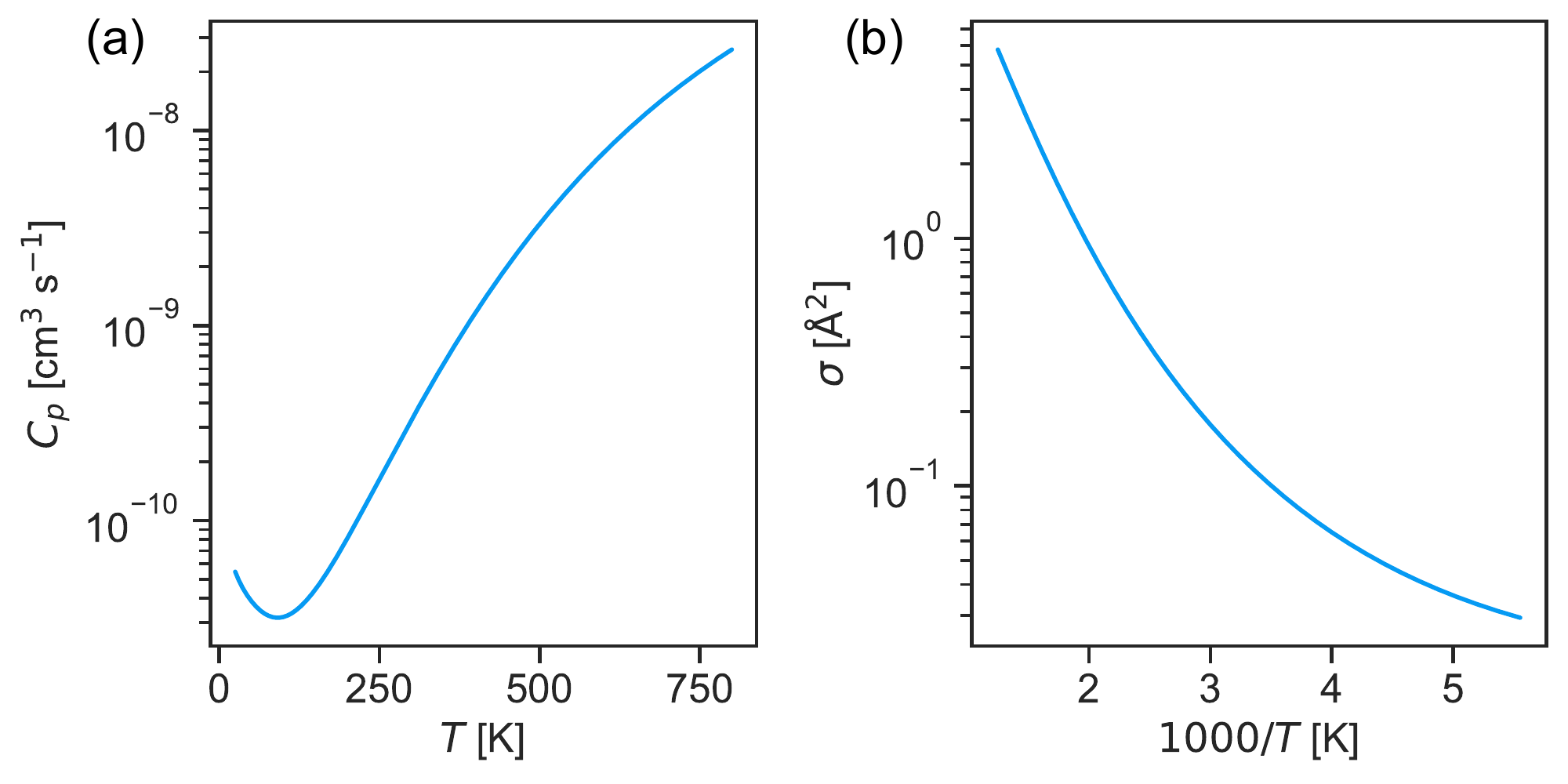}
    \caption{\label{fig:cap_coeff}
        The nonradiative (a) capture coefficient $C_p$ and (b) capture cross section $\sigma$ for the C$_{\rm N}$ defect in GaN computed with the \texttt{Nonrad} code.
    }
\end{figure}

With the input parameters extracted above, we are now ready to compute the capture coefficient.
We need to specify the volume of the supercell, which is 1100 \AA$^3$, and the configurational degeneracy of the final state $g$, which takes a value of 4 here because there are 4 (approximately) equivalent bond directions the defect state may relax to.
The \verb|get_C| function returns the \textit{unscaled} capture coefficient, i.e., we still need to determine $f$ in Eq.~\ref{eq:cap_coeff}.
For this defect system, the hole is captured by the negatively charged defect, so there will be a long-range Coulombic attraction that needs to be accounted for with the Sommerfeld parameter [contribution (i) of the scaling function].
If we had calculated the electron-phonon coupling in the negative charge state (point 2), we would also need to account for the perturbation of the band-edge wavefunction by the charged defect here [contribution (ii) of the scaling function].
The resulting hole capture coefficient $C_p$ and cross section $\sigma$ are shown in Fig.~\ref{fig:cap_coeff}a and Fig.~\ref{fig:cap_coeff}b, respectively.
The capture cross section $\sigma = C_p / \braket{v}$, where $\braket{v}$ is the average thermal velocity, may be obtained using the \verb|thermal_velocity| function provided by \texttt{Nonrad}.

\section{Summary}
\label{sec:conclusion}

We have presented the \texttt{Nonrad} code~\cite{nonrad} for the evaluation of nonradiative capture coefficients from first principles.
\texttt{Nonrad} implements the methodology developed by Alkauskas {\it et al.}~\cite{alkauskas_first-principles_2014}, which provides a quantum-mechanical description of the process of carrier capture at point defects.
We also presented a straightforward and tractable method for computing electron-phonon coupling to linear order within the PAW formalism.
Furthermore, we have shown that the process of replacing the Dirac delta functions by Gaussians can introduce significant errors;
our proposed interpolation scheme provides reliable results, even at low temperatures.
We also demonstrated that the analytic form of the Sommerfeld parameter may be insufficiently accurate: instead, we implement direct numerical integration to avoid troublesome approximations.
The \texttt{Nonrad} code provides a series of utilities, which facilitate the process of evaluating nonradiative capture coefficients.
This work will enable researchers to perform reliable studies of nonradiative capture in a range of materials.

\section*{Acknowledgements}
This work was supported by the U.S. Department of Energy, Office of Science, Basic Energy Sciences, under Award No. DE-SC0010689.
The Flatiron Institute is a division of the Simons Foundation.
Computational resources were provided by the National Energy Research Scientific Computing Center, a DOE Office of Science User Facility supported by the Office of Science of the U.S. Department of Energy under Contract No. DE-AC02-05CH11231.

\bibliographystyle{elsarticle-num}

\end{document}